\documentclass[english]{elsarticle}
\usepackage[T2A,T2A,T2A,T2A,T1]{fontenc}
\usepackage[latin9]{inputenc}
\usepackage{float}
\usepackage{units}
\usepackage{textcomp}
\usepackage{amsmath}
\usepackage{graphicx}
\usepackage{multirow}
\usepackage{multicol}
\usepackage{array,longtable}
\usepackage{babel}
\makeatletter

\journal{Canadian Journal of Physics}

\makeatother

\begin{document}
\begin{frontmatter}{}

\title{First-principle Calculations of Electron-Phonon Interactions in  $A^{II}B^{IV}C^{V}_2$ Crystals}

\author{V.G.Tyuterev}

\ead{valtyut00@mail.ru}

\address{Tomsk State Pedagogical University, Kievskaia st.,60,Tomsk, Russia}

\begin{abstract}
$Ab-initio$ probabilities of phonon-assisted intervalley scattering of electrons in the conduction bands of ternary chalcopyrite compounds  $ZnSiP_2$ and $ZnGeP_2$ between the  central $\Gamma$ minima and the lowest lateral minima (valleys) at $T$ and $N$ points have been calculated using the density functional theory. The equilibrium parameters of crystal structures, spectra of electrons and phonons are calculated self-consistently and are in fairly good agreement with the experiment and available theoretical calculations.  The electron-phonon coupling constants with short-wave (inter-valley) phonons in the chalcopyrite phosphides are close to their values in $Si$, $Ge$ and in binary analog $GaP$.
\end{abstract}
\end{frontmatter}{}

\section{Introduction.}\label{Intro}
Ternary semiconducting compounds $A^{II}B^{IV}C^{V}_2$ with chalcopyrite structure have long been intensively studied in connection with the search for an alternative to traditional working materials of semiconductor electronics from groups $IV$ and
$A^{III}B^{V}$  \cite{Shay,Alvarez1974,Poplavnoi_3}. Having the anisotropy of optical properties and rather large values of nonlinear optical coefficients, $A^{II}B^{IV}C^{V}_2$ chalcopyrite compounds proved to be promising candidates for technological applications as quantum generators in the infrared and optical parametric generators \cite{Vodopyanov}, light-emitting diodes in the visible and infrared ranges \cite{Patskun,Wandel}, infrared detectors \cite{Xiao}. The interest in the study of $A^{II}B^{IV}C^{V}_2 $ compounds  has increased in connection with the discovery of ferromagnetism at room temperature \cite{Gehlhoff}, as this creates potential for spintronic applications \cite{Medvedkin}. The topological insulators in the chalcopyrite structure were predicted recently \cite{Chen}.

The interest in chalcopyrites is supported due to the search of alternative energy sources.
Compounds of the type $ ZnXPn_2 $ ($X:Si,Ge$ and $Sn$;$Pn:P$ and $As$) are studying as thermoelectric materials with a high figure of merit. \cite{Neumann,Sreeparvath}. In recent years, the $ A^{II}B^{IV}C^{V}_2 $ compounds have come to the fore as materials for the solar energy transformation. The width of the band gap is in the range 1.0-2.4 eV, that in the combination with sufficiently large absorption coefficient and the stability of structure makes them promising candidates for the use in photovoltaics
\cite{Giles,Schilfgaarde_1}. Stable $A^{II}B^{IV}C^{V}_2$ compounds having a good match of the crystal structures with those of silicon, germanium or gallium arsenide are potential candidates for photovoltaic devices and can replace the more expensive $A^{I}B^{III}C^{VI}_2$ in photovoltaic applications. $ZnSiP_2$ due to the close match with the $Si$ lattice constant is of particular interest for the so-called tandem solar cells, that significantly improve the efficiency of traditional silicon solar cells \cite{Martinez}.

The investigation of electric and thermal transport in semiconductors requires the study of the charge carriers' scattering being the basic channel for momentum and energy  relaxation \cite{WangObuhTyutPRB_Thermo2011}. Scattering of electrons by long-wave phonons in chalcopyrites earlier was studied in our paper \cite{BorKarTyut1982_ang}.  In the presence of closely spaced energy minima in the conduction band the intervalley scattering (scattering by short-wavelength phonons) is also  important, especially for the study of energy relaxation at high excitation levels \cite{TyutZhukEcheniqChulk2015}.

Method of first-principle calculation for the probabilities of electron-phonon scattering in semiconductors first have been proposed in \cite{SjakTyutVastPRL,SjakTyutVastPRb,ObukGe}. In the current paper we study the intervalley electron-phonon scattering in $ZnSiP_2$ and $ZnGeP_2$ crystals with chalcopyrite structure.  Section \ref{method} describes the calculation procedure. In Sec.\ref{structura}, the crystal structure parameters of $ZnSiP_2$ and $ZnGeP_2$ were calculated. In  Sec.\ref{Bands}, electronic spectra are calculated. Sec.\ref{phonons} is devoted to the calculation of vibrational spectra. In Sec.\ref{Mejdolin} the intervalley scattering parameters (deformation potentials, hereafter DP) for the most significant electron-phonon transitions in the conduction band of $ZnSiP_2$ and $ZnGeP_2$ are calculated. We discuss the results in Conclusion.

\section{Calculation method.}\label{method}

The probability of electron scattering by a phonon with an arbitrary wavelength is determined as the square of modulus of the matrix element by the electron's wave functions  on of the perturbation of the crystalline potential created by this phonon. For a realistic, self-consistent calculation of the vibrational states, it is fundamentally important to begin with the finding of the equilibrium parameters of the crystal structure by the first principles. The single-particle energies and wave functions of electrons then should be found self-consistently with the optimized parameters of the structure.

We performed these program for the crystals $ZnSiP_2$ and $ZnGeP_2$  within the pseudopotential method in the framework of the theory of the electron density functional $DFT$. The software package $Quantum$ $Espresso$ \cite{ESPRESSO} have been used.
Hard-core pseudopotentials Zn.pz-van-ak.UPF,  Si.pz-vbc.UPF,  Ge.pz-bhs.UPF, P.pz-bhs.UPF, respectively for zinc, silicon, germanium and phosphorus with Perdue-Zunger exchange-correlation potential in the generalized gradient approximation $GGA$  are taken from $Quantum$ $Espresso$ website $http://www.quantum-espresso.org/pseudopotentials$. The one-particle Kohn-Sham wave functions were expanded into plane waves series with the kinetic energy cutoff $ecutwfc=60eV$. Integrals by the Brillouin zone were calculated by Monkhorst-Pack special points method with a partition (444 111) for the chalcopyrite structure.

\section{Optimisation of crystal structure.}\label{structura}

\begin{figure}[h]
\begin{center}
	\begin{minipage}[h]{0.40\linewidth}
		\includegraphics[width=1.0\linewidth]{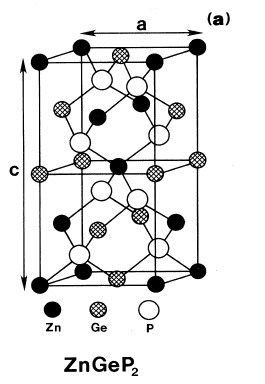}
	\end{minipage}
	\begin{minipage}[h]{0.59\linewidth}
		\includegraphics[width=1.0\linewidth]{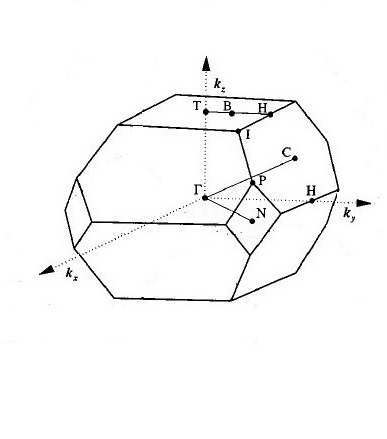}
	\end{minipage}
	\caption{\small{Structure and Brillouin zone of chalcopyrite}}
	\label{fig:Cell}
\end{center}
\end{figure}

Chalcopyrite structure is a body-centered tetragonal lattice with space group $D^{12}_{2d}$
($I\overline{4}2d$)). $A^{II}B^{IV}C^{V}_2$ crystal is close to the sphalerite $III-V$ (zinc blende) lattice but with the ordered alternate positions of $A^{II}$ and $B^{IV}$ in cations' sublattice. This is accompanied by a slight distortion of the tetrahedral coordination of $C^V$ anions that makes the unit cell tetragonal with the tetragonal $c$ axis being roughly twice the basal plane lattice constant $a$. Parameter $c/a$ defines the tetragonal contraction of the structure. So-called tetrahedral distortion is represented by the internal structural parameter $u$, which determines the shift of the anion $C^V$ off the ideal tetrahedral position with $u_{ideal}=0.25$. The elementary cell of chalcopyrite structure contains four molecules per unit cell Fig.\ref{fig:Cell}.

The basis vectors in the units of the basal plane lattice constant $a$ are chosen as follows:
$\mathbf{a}\left(1\right)=a(1/2,-1/2, \gamma)$, $\mathbf{a}\left(2\right)=a(1/2,1/2, \gamma)$, $\mathbf{a}\left(3\right)=a(-1/2,-1/2,\gamma)$, here $\gamma$=$c/(2a)$.

Atomic positions in the unit cell of chalcopyrite are:
$A^{II}$$(1,0,0)$,
$(1,1/2,\gamma/2)$;
$B^{IV}$$(1/2,0,\gamma/2)$,
$(1/2,1/2,0)$;
$C^{V}$$(1/4,1/2-u,3\gamma/4)$,
$(u,3/4,\gamma/4)$,
$(-u,1/4,\gamma/4)$,
$(3/4,1/2+u,3\gamma/4)$.

The equilibrium crystal structure was determined in the framework of the $DFT$ method using the $BFGS$
 optimisation technique \cite{Fischer} in the $Quantum$ $Espresso$ \cite{ESPRESSO} implementation, which provides a fast method of finding the lowest energy structures. The tolerance for the geometry optimizing was set as the difference in the total energy within
 $10^{-8}eV$, the maximum value of the force acting on the atom at the equilibrium position was limited to $F_i\leq0.001Ry/Bohr$, the maximum value of the mechanical stress is $P\leq0.2kBar$.

  \begin{table*}[ht]

\footnotesize

	\begin{center}
		\caption{\small{Parameters of crystal structure of $ZnSiP_2$ and $ZnGeP_2$ }} \label{Ts:Structura}
 \begin{tabular}{|c|c|c|c|c|c|c|}
 	\hline
     \multicolumn{1}{|c}{} &\multicolumn{3}{|c|}{} & \multicolumn{3}{c|}{} \\
 	\multicolumn{1}{|c}{} &\multicolumn{3}{|c|}{\small{$ZnSiP_2$}} & \multicolumn{3}{c|}{\small{$ZnGeP_2$}} \\
    \multicolumn{1}{|c}{} &\multicolumn{3}{|c|}{} & \multicolumn{3}{c|}{} \\
 	\hline
 	\small{Structure}&\small{Our}&\small{Other}&\small{Exper.}&\small{Our}&\small{Other}&\small{Exper.}\\
 \small{parameter}&\small{calculation}&\small{theories}&\cite{MacKinnon}&\small{calculation}&\small{theories}     & \cite{MacKinnon,Lind}\\
  & &\cite{Sreeparvath,Martinez,Sharma,Shaposhnikov,Pena,Arab} && & \cite{Sreeparvath,Limpijumnong_1,Jaffe,Tripathy} &\\
   	\hline
 $a$($\mathring{A}$)&5.292&5.35$\div $5.5062& 5.400& 5.3617 &5.396$\div $5.52 & 5.459$\div $5.465\\
      	\hline
  $c/a$ &1.958&1.932$\div $1.946 & 1.933 & 1.976  &1.958$\div $1.986 & 1.958$\div $1.970\\
    	\hline
 	  $u$&0.2622& 0.268$\div $0.26948 &0.2691& 0.2507 &0.25$\div $0.254& 0.25816$\div $0.267\\
   	\hline
 	
 \end{tabular}
   \end{center}
\end{table*}

Deviations
from the experimental values of the parameters do not exceed 3\%
(Tab.\ref{Ts:Structura}) and are most significant for the deformation parameter of tetrahedra $u$ in $ZnGeP_2$.

The values of lattice parameters obtained in other theoretical studies differ from the experimental values in the range less than
1\%,
but the scatter in various calculations is rather significant (Tab.\ref{Ts:Structura}), which is not surprising, since the results of structure optimization in $DFT$ depend on the choice of pseudopotentials and the form of the exchange-correlation potential.

Our ultimate goal is to calculate the parameters of the electron-phonon coupling, so the optimization of the structure, the calculation of the states of both electrons and phonons should be carried out within the framework of a unified self-consistent approach. Worth note that we obtain the correct topological structure of the electronic bands (sf Sec.\ref{Bands}) and simultaneously reach an acceptable good agreement with experiment on the structural parameters and phonon frequencies (sf Sec.\ref{phonons}).

\section{Band structure of  $ZnSiP_2$ and $ZnGeP_2$.} \label{Bands}

Calculation of electron band structure in $ZnSiP_2$ and $ZnGeP_2$ was carried out within the electron density functional  theory  $DFT$.
 We used the software package $ Quantum$ $Espresso $ \cite{ESPRESSO}, with the same as in Sec.\ref{structura} hard core pseudopotentials
taken from $ Quantum$ $Espresso$  website.
 Tab.\ref{Ts:Levels} give our calculated data, as well as theoretical calculations, known from the literature.

\begin{table*}[ht]
\footnotesize
	\begin{center}
		\caption{\small{Energy levels in the symmetry points of Brillouin zone. The top of the valence band $\Gamma_{4v}$ was chosen as the origin.
Data of other calculations are taken from \cite{Limpijumnong_1}$^{a}$,\cite{Jaffe}$^{b}$,\cite{Chiker}$^{c}$, $^{(*)}$- corrected in the scissor approximation.}}\label{Ts:Levels}
 \begin{tabular}{|c|c|c|c|c|}
 	\hline
 	\multicolumn{1}{|c|}{Level} & \multicolumn{2}{c|}{\small{$ZnSiP_2$ }} & \multicolumn{2}{c|}{\small{$ZnGeP_2$}} \\
 	\hline
 	            &\small{Our}    & \small{Other }      &\small{Our}        & \small{Other }  \\
                &\small{Calculation} &\small{theories}        & \small{ calculation}   &\small{theories}  \\
 	\hline
 	 $\Gamma_{1c}$ &    2.71           & 1.134$^{c}$   &    1.76               & 1.24$^{a}$,2.49$^{a*}$,1.19$^{b}$  \\
   	\hline
  $\Gamma_{2c}$ & 1.71          &2.37$^{b}$,1.365$^{c}$ & 1.52 & 1.54$^{a}$,2.31$^{a*}$,1.53$^{b}$,1.079$^{c}$\\
   	\hline
  $\Gamma_{3c}$ & 1.09          & 1.11$^{b}$,1.732$^{c}$ & 1.16 &1.27$^{a}$,2.67$^{a*}$,1.09 $^{b}$, 1.397$^{c}$\\
   	\hline
  $T_{1c}+T_{2c}$ & 1.27        & 1.40$^{b}$,2.609$^{c}$ & 1.24 &1.20$^{a}$,2.24$^{a*}$,1.23$^{b}$,1.445$^{c}$ \\
   	\hline
  $T_{5c}$      & 1.51          & 1.67$^{b}$    & 1.17 & 1.2$^{b}$ \\
   	\hline
  $N_{1c}$     & 2.00          & 1.91$^{b}$,1.958$^{c}$   & 1.35 &1.14$^{a}$,2.31$^{a*}$,1.171$^{c}$ \\
     \hline
   $\Gamma_{4v}$ &  0            &  0                  & 0    & 0 \\
         \hline
  $\Gamma_{5v}$  &   -0.12      & -1.135$^{c}$, -0.30$^{b}$  & -0.07    & -0.33 $^{b}$  \\
    \hline
  $T_{3v}+T_{4v}$ & -1.26       & -1.25$^{b}$   & -1.45 & -1.50 $^{b}$ \\
    	\hline
   $N_{1v}$    & -0.58             & -0.67$^{b}$    & -0.78     & -0.86 $^{b}$ \\
   	\hline
 \end{tabular}
   \end{center}
\end{table*}

Our values of forbidden gaps $E_g$: $1.09eV$ in $ZnSiP_2$ and $1.15eV$ in $ZnGeP_2$ -  are in fairly good agreement with the previous theoretical $DFT$ calculations
 \cite{Shaposhnikov,Arab,Chiker,Ulla},
 however their width is underestimated in comparison with experiment \cite{Shay}. This is a well-known problem for the Kohn-Sham states in semiconductors and insulators  in $DFT$. It is solved in many-particle methods
 but at the cost of a significant increase in the amount and time of calculations. In any case, the issue of gap width as itself is not relevant to our aim since here we are interested in the investigation of electron-phonon scattering only inside the conduction band.

 As for conduction bands, the literature data concerning the classification of states and, in particular, the symmetry assignment of the conduction band minimum ($ CBM $), are contradictory. The point is that, despite its structural similarity, various compounds
  $ A ^ {II} B ^ {IV} C ^ {V} _2 $ can demonstrate either direct, pseudo-direct, or indirect type of optical absorption.

While the maximum of the valence band ($ VBM $) with symmetry $ \Gamma_ {4v} $ is always at the centre of BZ, the conduction band minimum
($ CBM $) in various compounds can come to be in the centre of BZ, having the symmetry $\Gamma_ {1c} $ (direct transition) or $ \Gamma_ {3c} $ (pseudo-direct transition),  as well as it may occur at BZ border (indirect transition). The 'pseudo-direct' band gap is nominally direct but the inter-band optical transition
$ \Gamma_ {4v} $-$ \Gamma_ {3c} $  is forbidden by symmetry in the first order of perturbation theory. Like the indirect transition, the pseudo-direct transition is associated with weak optical absorption near the band edge.
Until now, there is no generally accepted opinion on the structure of conduction bands in compounds $ZnSiP_2$ and $ZnGeP_2$.

 Chiker et al \cite{Chiker}, Rashkeev et al. \cite{Rashkeev}
 showed that $ZnSiP_2$ is a pseudo-direct material with $CBM$ at $\Gamma$ - point, unlike Shay et al. \cite{Shay}, who predicted a direct forbidden gap for this compound. For $ZnGeP_2$  De Alvarez et al. \cite{Alvarez1974} predicted a direct forbidden gap ($\Gamma_{4v}$ - $\Gamma_{1c}$) using the empirical pseudopotential method.

 Rashkeev et al. \cite{Rashkeev}
  have discovered the indirect  nature of the band gap in $ZnGeP_2$ with the conduction band minimum close to the point $N$.

Shaposhnikov et al \cite{Shaposhnikov} investigated band structures of chalcopyrites by various approaches
$ZnSiP_2$ in their calculation turns out to be a direct-band semiconductor. $ZnGeP_2$ appears to be the pseudo-direct with
$CBM$ at $\Gamma$ - point, whereas the values of minima in $\Gamma$ and $N$ -points have practically equal energy values.

Chiker et al. \cite{Chiker} have found $CBM$ in the $T$ -point predicting that $ZnGeP_2$ is the indirect band-gap material.

Limpijumnong et al. \cite{Limpijumnong_1} have found in $ZnGeP_2$ several closely lying $CBM$, from which the lowest-lying is at low-symmetry point of Brillouin zone.

 \begin{figure}[h]
\begin{center}
	\begin{minipage}[h]{0.49\linewidth}
		\includegraphics[width=1\linewidth]{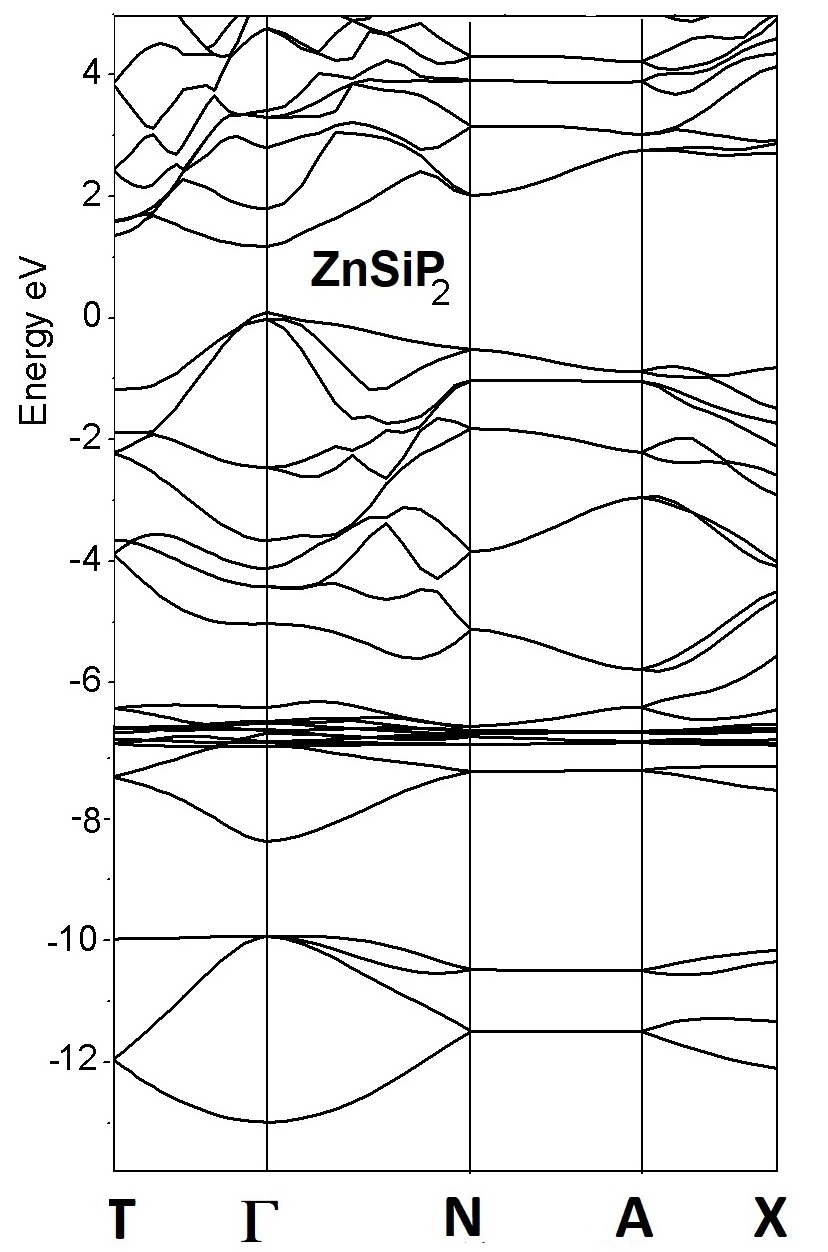}
	\end{minipage}
	\begin{minipage}[h]{0.495\linewidth}
		\includegraphics[width=1\linewidth]{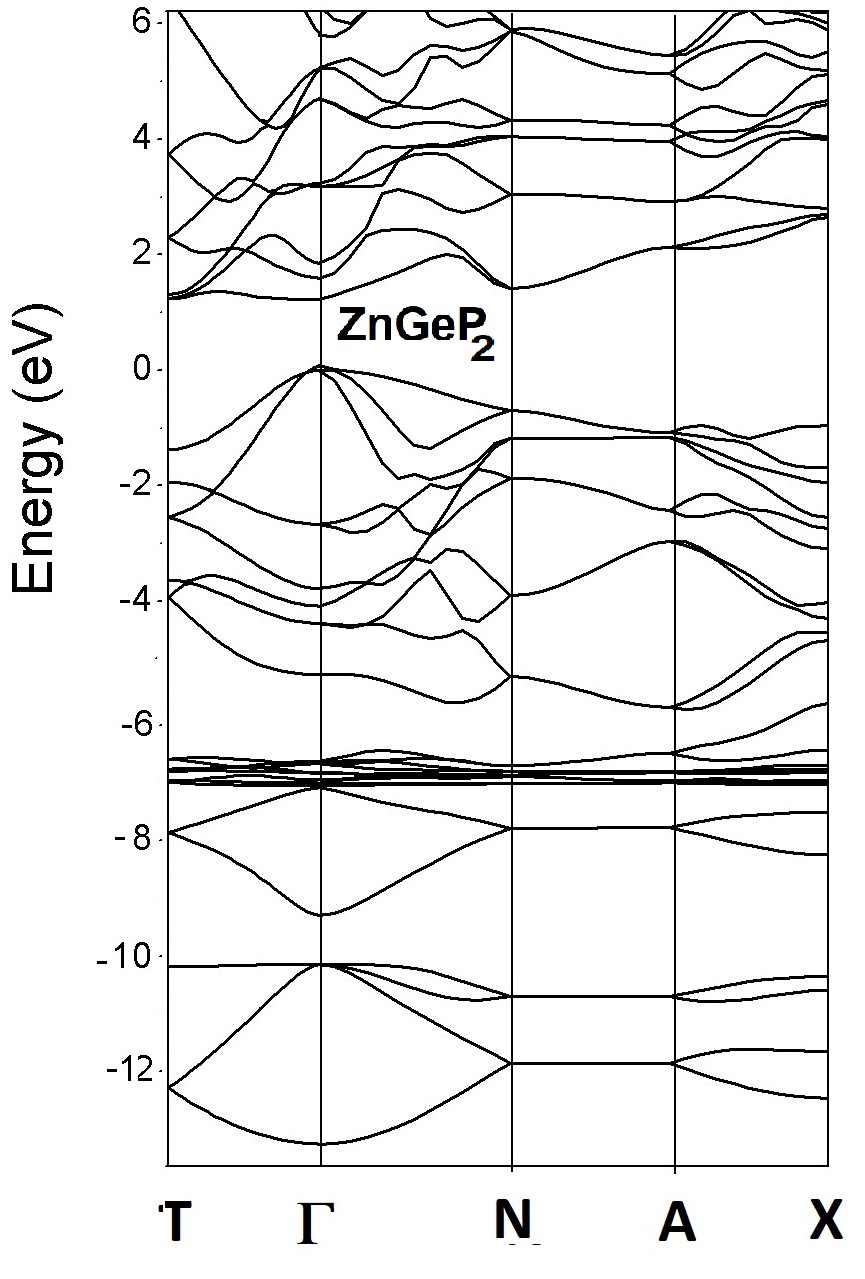}
	\end{minipage}
	\caption{\small{Calculated band spectra of $ZnSiP_2$ and $ZnGeP_2$}}
	\label{fig:bands}
\end{center}
\end{figure}
Our calculated energy spectra of  $ZnSiP_2$ and $ZnGeP_2$ Fig.\ref{fig:bands} reveal the topological structure which is in a good overall agrement with other calculations

 \cite{Shaposhnikov,Chiker,Ulla,Rashkeev,Limpijumnong_1}.
 The valence band maximum ($VBM$) and the conduction band minimum  ($CBM$) are located at the center of Brillouin zone $\Gamma$ in both compounds. According to our calculations, both compounds have a minimum forbidden gap at the point $\Gamma$ with $\Gamma_{3c}$ representing the minimum of the conduction band, hence both compounds are pseudo-direct. Interestingly, in $ZnGeP_2$, the conduction band states $T_{5c}$ and $T_{1c} + T_{3c}$ at the edge of the Brillouin zone at point $T$  turned out to be very close to $\Gamma_{3c}$. Hence, $ZnGeP_2$ reveals a proximity to indirect behavior resembling those in
 \cite{Chiker} and \cite{Limpijumnong_1}.

\section{Phonon spectra of $ZnSiP_2$ and $ZnGeP_2$.} \label{phonons}

Experimental data on the phonon spectra of $ZnSiP_2$ and $ZnGeP_2$ relate only to the long-wave phonons and are known from optical experiments on the infrared absorption and Raman scattering of light \cite{Pena,Ohendoprf_I_II,Shportko,Landolt_1}. Theoretical calculations were carried out earlier in the framework of phenomenological theories  in  \cite{Ohendoprf_I_II,PoplTjuJdePhys}. According to the group-theoretical study,  the vibrational representation in the center of Brillouin zone is of the form:
 $\Gamma_1(A_1)$+ 2$\Gamma_2(A_2)$+3$\Gamma_3(B_1)$+4$\Gamma_4(B_2)$+7$\Gamma_5(E)$.
Phonons with symmetry  $\Gamma_1$, $\Gamma_3$, $\Gamma_4$, $\Gamma_5$ are active in Raman scattering of light. $\Gamma_4$ and $\Gamma_5$ are active also for the infrared absorption of light. Phonons  $\Gamma_2$ are not active in optical experiments. Chalcopyrite compounds are the polar crystals having the ion-covalent character of the chemical bond. In the long-wavelength spectrum, the oscillations related to the representations of $\Gamma_4(B_2)$ and $\Gamma_5(E)$ are split into longitudinal (L) and transversal (T)  components.

Phonon spectra in some of $A^{II}B^{IV}C^{V}_2$ crystals were calculated in the frames of  $DFT$ in the works \cite{LazewskiParl} and showed a good agreement with  experiment. The ab-initio calculation of long-wavelength spectrum for $ZnSiP_2$ was performed in \cite{Bhadram}. Calculations of phonon spectra of $ZnGeP_2$ from the first principles are absent in the available literature. Our calculations were performed using  the density functional perturbation theory  ($DFPT$) proposed in the article \cite{BaroniGirDal} and implemented in $Quantum Espresso$ \cite{ESPRESSO} package.

The calculations were performed with the values of the structure parameters obtained in
Sec.\ref{structura} and the same pseudopotentials. Integration in the Brillouin zone was performed using a set of special Monhorst-Pack points [444, 111] and energy cutoff $ Ecut = 60 eV $.

 \begin{figure}[h]
 \begin{center}
	\begin{minipage}[h]{0.485\linewidth}
		\includegraphics[width=1\linewidth]{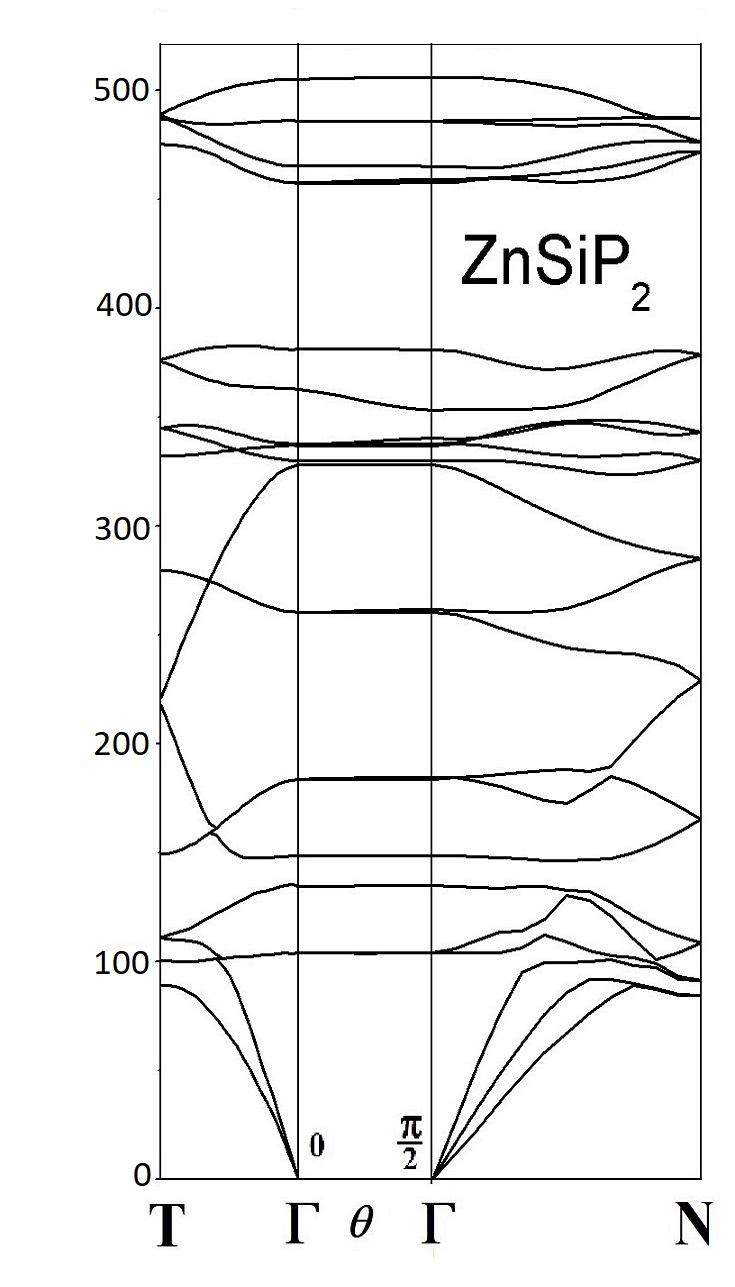}
	\end{minipage}
	\begin{minipage}[h]{0.495\linewidth}
		\includegraphics[width=1\linewidth]{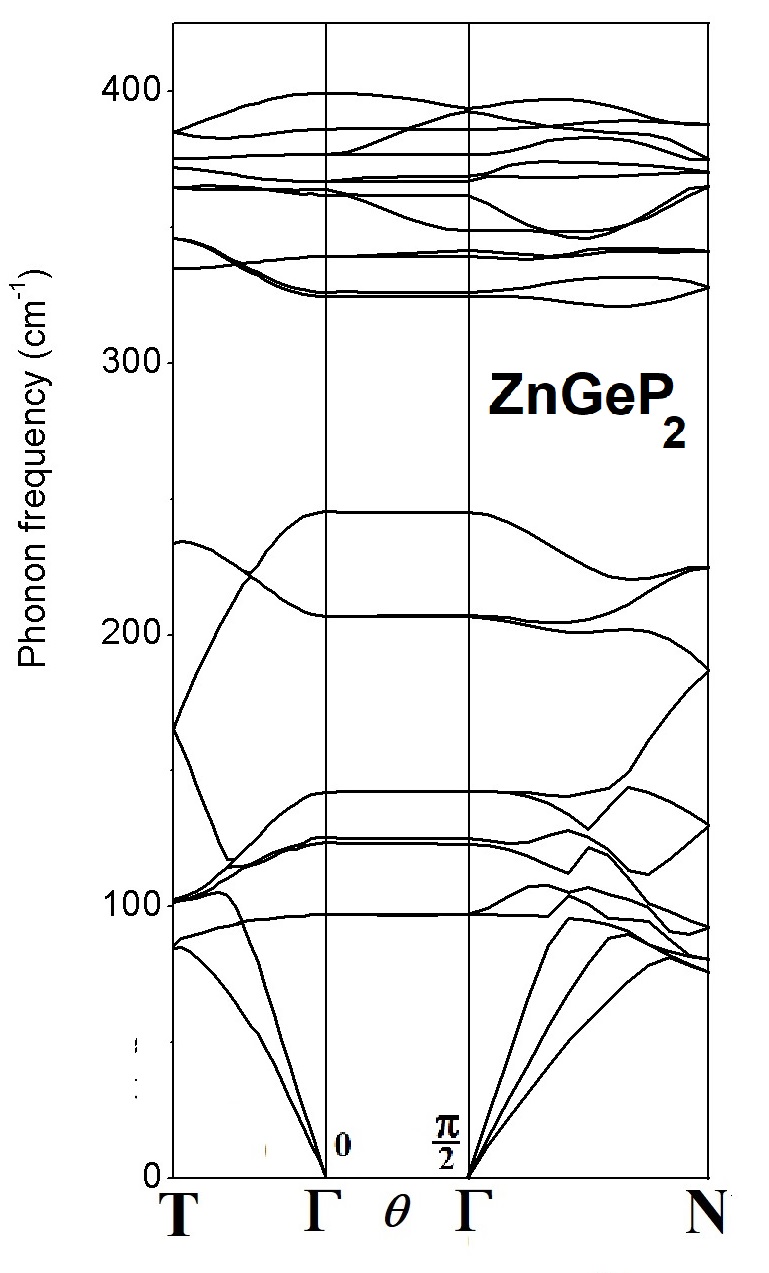}
	\end{minipage}
	\caption{\small{Calculated phonon spectra of $ZnSiP_2$ and $ZnGeP_2$ crystals.  Central part dispayes the angle dispersion of frequencies in the long-wavelength limit $q \rightarrow 0$ depending on the angle $\theta$ between the direction of the phonon propagation and the tetragonal axis $c$.}}
	\label{fig:phonons}
\end{center}
\end{figure}
Frequencies of long-wavelength phonons, calculated with $DFPT$, are presented in
Tab.\ref{Ts:phonZSGP} in comparison with experiment. For $ZnGeP_2$ calculations from \cite{Bhadram} are also given.
\begin{table*}[ht]
\footnotesize
	\begin{center}
		\caption{\small{Longitudinal/transversal long-wavelength phonons in $ZnSiP_2 $ and $ZnGeP_2$}(units $cm^{-1}$)} \label{Ts:phonZSGP}
 \begin{tabular}{|c|c|c|c|c|c|c|c|}
 \hline
 \multicolumn{1}{|c}{}&\multicolumn{4}{|c}{$ZnSiP_2$}&\multicolumn{3}{|c|}{$ZnGeP_2$}\\
 \cline{2-8}
 \multicolumn{1}{|c}{Symmetry}&\multicolumn{2}{|c}{Theory}&\multicolumn{2}{|c}{Experiment}&\multicolumn{1}{|c}{Our}
 &\multicolumn{2}{|c|}{Experiment}\\
  \cline{2-5}
   \cline{7-8}
 \footnotesize{}   &\footnotesize{Our}&{Ref.\cite{Bhadram}}&{Ref.\cite{Landolt_1}}&{Ref.\cite{Pena}}&{ theory}&\footnotesize{Ref.\cite{Landolt_1}}
 &{Ref.\cite{Shportko}}\\
 	\hline
 \footnotesize{}           &\footnotesize{505/486}&521/497&\footnotesize{518/494}&\footnotesize{518/494}&\footnotesize{394/377}&\footnotesize{404/387}&-/386\\
 \footnotesize{}          &\footnotesize{459/458}&465/-&\footnotesize{464/461}&\footnotesize{-/462}&\footnotesize{369/367}&\footnotesize{376/369}&-/368\\
 \footnotesize{$\Gamma_5(E)$}&\footnotesize{340/338}&-&\footnotesize{327/321}&\footnotesize{-}&\footnotesize{342/339}
 &\footnotesize{331/329}&-/327\\
 \footnotesize{}           &\footnotesize{262/260}&-/267&\footnotesize{264/246}&\footnotesize{-/263}&\footnotesize{207/207}&\footnotesize{204/202}&-/202\\
 \footnotesize{}           &\footnotesize{185/184}&-/187&\footnotesize{187/185}&\footnotesize{-/183}&\footnotesize{143/143}&\footnotesize{-/142}&-/142\\
 \footnotesize{}          &\footnotesize{104/104}&-/105&\footnotesize{102/102}&\footnotesize{-/101}&\footnotesize{97/97}&\footnotesize{-/96}&-\\
  	\hline
             &\footnotesize{505/486}&521/497&\footnotesize{518/494}&\footnotesize{518/494}&\footnotesize{399/393}
             &\footnotesize{408/399}&-/400\\
 \footnotesize{$\Gamma_4(B_2)$} &\footnotesize{362/353}&355/347&\footnotesize{359/343}&\footnotesize{ - }&\footnotesize{362/349}&\footnotesize{359/344}&-/342\\
          &\footnotesize{149/149}&-/147&\footnotesize{145/140}&\footnotesize{-}&\footnotesize{123/123}&\footnotesize{-}&-/116\\
\hline
            &\footnotesize{465}&{-}&\footnotesize{466}&\footnotesize{461}&\footnotesize{386}&\footnotesize{389}&-\\
 \footnotesize{$\Gamma_3(B_1)$} &\footnotesize{328}&339&\footnotesize{335}&\footnotesize{336}&\footnotesize{245}&\footnotesize{247}&-\\
          &\footnotesize{135}&132&\footnotesize{130}&\footnotesize{128}&\footnotesize{125}&\footnotesize{120}&-\\
 \hline
  \footnotesize{$\Gamma_2(A_2)$} &\footnotesize{381}&{-}&\footnotesize{ - }&\footnotesize{-}&\footnotesize{362}&\footnotesize{-}&-\\
          &\footnotesize{330}&\footnotesize{ - }&{-}&\footnotesize{-}&\footnotesize{324}&\footnotesize{-}&-\\
 \hline
 \footnotesize{$\Gamma_1(A_1)$} &\footnotesize{338}&337&\footnotesize{337}&\footnotesize{337}&\footnotesize{326}&\footnotesize{328}&-\\
 \hline
  \end{tabular}
   \end{center}
\end{table*}

As it can be seen from the Tab.\ref{Ts:phonZSGP}, the long-wavelength frequencies in our \emph{ab-initio} calculations are in a fairly good agreement with experiment. For the calculating of the inter-valley scattering the short-wavelength  phonons are nesessary, therefore the spectra at the boundaries of the Brillouin zones also should be calculated. The phonon spectra along two high-symmetry directions of BZ are shown in Fig.{\ref{fig:phonons}. Worth note that they are in qualitative agreement with the results of our early phenomenological calculations
\cite{PoplTjuJdePhys}. Note also that these are the first \emph{ab-initio} calculations of the vibrational spectra in
$ ZnSiP_2 $ and $ ZnGeP_2 $.

\section{Intervalley scattering.} \label{Mejdolin}
The probability of an electron transition from the initial Bloch state
$\Psi_{n\mathbf{k}}$ ($n$-number of the band, $\mathbf{k}$-wave vector) to the
final state $\Psi _{n^{\prime }\mathbf{k}\pm \mathbf{q}}$ under the influence of
the crystal potential perturbation $V_{\mathbf{q}^{\prime
}}^{\lambda }$ caused by the phonon of the $\lambda^{th} $ branch with the wave
vector $\mathbf{q}$ and the frequency $\omega^\lambda _\mathbf{q}$ is defined as follows:

\begin{eqnarray}
W_{n\mathbf{k},n^{\prime }\mathbf{k}\pm \mathbf{q}}=\frac{2\pi }{\hbar }%
\left\vert \langle \Psi _{n\mathbf{k}}\left\vert
V^\lambda_{ \mathbf{q}}\right\vert\Psi _{n^{\prime }\mathbf{k}\pm \mathbf{q}}\rangle
\right\vert ^{2}\times &\label{eq:transition_probability}\\
\left( N(\omega^\lambda _{\mathbf{q}},T)+\frac{1}{2}\mp \frac{1}{2}%
\right) \delta \left( E_{n\mathbf{k}}\pm \hbar \omega^\lambda _{\mathbf{q}%
}-E_{n^{\prime }\mathbf{k}\pm \mathbf{q}}\right)&\nonumber
\end{eqnarray}

In this expression, the delta function expresses the energy conservation
law, $N\left( \omega _{\mathbf{q}}^{\lambda },T\right) $ are the equilibrium
phonon occupation numbers, the upper sign in $W_{n\mathbf{k},n^{\prime }\mathbf{k}\pm \mathbf{q}}$
 corresponds to absorption, the lower sign to the
emission of a phonon.

The perturbation of the crystal potential caused by the field of atomic
displacements is a superposition of contributions from different types of
phonons. The contribution from a phonon with the wave vector $\mathbf{q}$,
belonging to the spectrum branch of the number $\lambda$, can be represented
in the form

\begin{equation}
V_{\lambda \mathbf{q}}=\sum\limits_{s\alpha }\sqrt{\frac{M}{m_{s}}}%
\sqrt{\frac{\hbar }{2\rho \Omega \omega _{\lambda \mathbf{q}}}}e_{\lambda \mathbf{q%
}}^{s\alpha }\delta V_{\mathbf{q}}^{s\alpha }\left( \mathbf{r}\right)
\label{unit_perturbation}
\end{equation}

Here, $\delta V_{\mathbf{q}}^{s\alpha }\left( \mathbf{r}\right)$
 is the
perturbation of crystal potential caused by a unit displacement of the
$s^{th}$ atom in the $\alpha$ direction involved into the propagation of the
phonon with the wave vector $\mathbf{q}$; $\omega^\lambda _{ q}$ and
$e^{s\alpha}_{\lambda \mathbf{q}}$ are respectively the frequency and
eigenvector of the phonon; $m_{s}$ is the mass of the atom, $M$ is the mass
of the unit cell; $\rho$, $\Omega$ are the mass' density and the volume of the
crystal

It is convenient to introduce a generalized deformation potential for phonon
with an arbitrary wavelength by the relation
(Grinyaev, Karavaev and Tyuterev \cite{GrinKarTyut89-ang}):

\begin{equation}
D^{nn^{\prime }\lambda }_{\mathbf{k},\mathbf{k^{\prime }}}=\sqrt{\frac{%
2\rho \Omega\omega^\lambda _{ \mathbf{k}-\mathbf{k^{\prime }}}}{\hbar }}\left\vert \langle \Psi _{n%
\mathbf{k}}\left\vert V^\lambda_{\mathbf{k}-\mathbf{k^{\prime }} }\right\vert \Psi
_{n^{\prime }\mathbf{k^{\prime }}}\rangle \right\vert
\label{eq:general_defpot}
\end{equation}

Deformation potential for scattering by short-wave phonon
$D^{nn^{\prime }\lambda }_{\mathbf{k},\mathbf{k^{\prime }}}$
 depends on the properties
of the electron and phonon subsystems through the wave functions of the
initial and final states, phonon eigenvector
 $e_{\lambda \mathbf{q}}^{s\alpha }$ and the perturbation of the atomic potential caused by
this oscillation.

The case when the electron is scattered from a vicinity of a minimum
(valleys) in the band spectrum to a vicinity of another minimum is the most
interesting from the point of view of application to the analysis of various
physical properties. This process is called as intervalley scattering. The
scheme of such processes in the conduction band $ZnSiP_2$ and $ZnGeP_2$ is
shown in Fig.\ref{fig:ZSP_trans}.

Let as investigate intervalley scattering from the band minimum at point $\mathbf{k=k}_{i}$
(hereafter bottom of $i^{th}$ valley)
into the vicinity $\mathbf{k}_{j}+\mathbf{\kappa }$
of another band minimum ( hereafter $j^{th}$ valley),considering
$E_{n\mathbf{k}_{i}}\geq E_{n^{\prime }\mathbf{k}_{j}+\mathbf{\kappa }}$,
 Here vector $\mathbf{\kappa }$ is counted up the position of the bottom of $j^{th}$ valley.
The phonon energy is small in comparison with electron energies:
$\hbar \omega _{\mathbf{k}_{j}-\mathbf{k}_{i}\pm \mathbf{\kappa }}^{\lambda
}\ll E_{n\mathbf{k}_{i}},E_{n^{\prime }\mathbf{k}_{j}\pm \mathbf{\kappa }}$
so we can neglect it in the argument of the $\delta $ -function
in Eq.(\ref{eq:transition_probability}) and so to consider the phonon emission and absorption in equal
footings.

In typical case when the deformation potentials and phonon frequencies
depend weakly on $\kappa $ in the neighborhood of the $j^{th}$ valley,
$D^{nn^{\prime}\lambda }_{\mathbf{k}_i,\mathbf{k}_j+\mathbf{\kappa }}
\approx D^{nn^{\prime}\lambda }_{\mathbf{k}_i,\mathbf{k}_j}$
  and
$\omega _{\mathbf{k}_{j}-\mathbf{k}_{i}\pm \mathbf{\kappa }}^{\lambda }\approx \omega
_{\mathbf{k}_{j}-\mathbf{k}_{i}}^{\lambda }$,
then the probability of the
electron transition from the bottom of the $i^{th}$ valley to the $j^{th}$
valley involving the all phonon spectra $\lambda ^{th}$ branch can be
written in the form:
\begin{equation}
W_{j}^{i}=R_{j}^{i}\left( T\right)G_{j}
\end{equation}
Here
\begin{equation}
R_{j}^{i}\left( T\right) =\frac{\pi }{\rho V}\sum\limits_{\lambda }\frac{%
\left\vert D^{nn^{\prime}\lambda }_{\mathbf{k}_i,\mathbf{k}_j}
\right\vert ^{2}}{\omega _{\mathbf{k}_{j}-\mathbf{k}_{i}}^{\lambda }}\left(
2N\left( \omega _{\mathbf{k}_{j}-\mathbf{k}_{i}}^{\lambda },T\right)
+1\right)\label{rates}
\end{equation}
is the temperature-dependent rate of $i \rightarrow j$ transition and
$G_{j}=\sum\limits_{\mathbf{\kappa }}\delta
\left( E_{n^{\prime }\mathbf{k}_{i}}-E_{n^{\prime }\mathbf{k}_{j}+\mathbf{%
\kappa }}\right)$
is the density of electronic states in the $j^{th}$ valley.

The use of the electron density  functional  perturbation theory
($ DFPT $) allows one to calculate the deformation potentials for electron-phonon
transitions between any points in the Brillouin zone. In the above-mentioned approximations,
it is sufficient to calculate the value of the deformation potential only for the
"virtual transition" between the bottoms of the valleys, avoiding massive
$\mathbf {\kappa} $-dependent calculations. Although formally the energy conservation law
is not satisfied, as it should be according to Eq.(\ref{eq:transition_probability}), the analysis of heat and charge transfer processes in this
 approach (the Conwell approximation)  proved its practical applicability in semiconductors
\cite{Ridley}.

.

A numerical method for calculating the probabilities of the interaction of
electrons with phonons for metals is realized in the Quantum Espresso \cite{ESPRESSO} software package and modified in \cite{SjakTyutVastPRL,SjakTyutVastPRb} for nonconducting crystals.

In the process of calculating the spectrum of lattice vibrations by the $DFPT $ method \cite{BaroniGirDal}, the perturbation $\delta \hat{V}^{s
\alpha}_\mathbf{q}$ of the the crystal potential, caused by the displacement
of the atom in the phonon vibrations, is calculated self-consistently .

The wave functions $|\Psi _{n\mathbf{k}}\left. {}\right\rangle $, $|\Psi
_{n^{\prime }\mathbf{k}+\mathbf{q}}\left. {}\right\rangle $ in the matrix
element for the electron-phonon transition (\ref{eq:transition_probability}%
), energies $E_{n\mathbf{k}},$ $E_{n^{\prime }\mathbf{k}+\mathbf{q}}$ and
the electron-phonon interaction operators $\ \delta \hat{V}_{\mathbf{q}%
}^{s\alpha }$, as well as the solution of the dynamical problem $\omega^\lambda
_{\mathbf{q}}$, and $e_{\mathbf{q}\lambda }^{s}$ for a phonon of an
arbitrary wavelength are calculated within the same self-consistent routine.
Thus, the $DFPT $ method allows one to determine in a self-consistent manner
and from the first principles the transition matrix element for any
electron-phonon scattering channel under consideration. This completely
determines the calculation of the deformation potential

Deformation potentials for intervalley transitions in chalcopyrite crystals were not calculated before.
Transitions between the central minimum in $\Gamma$ of the conduction band and the closely spaced energy minima at the points
$T$ and $N$ deserve the most attention for the analysis of relaxation processes.

\begin{figure}[h]
\begin{center}
	\begin{minipage}[h]{0.490\linewidth}
		\includegraphics[width=1\linewidth]{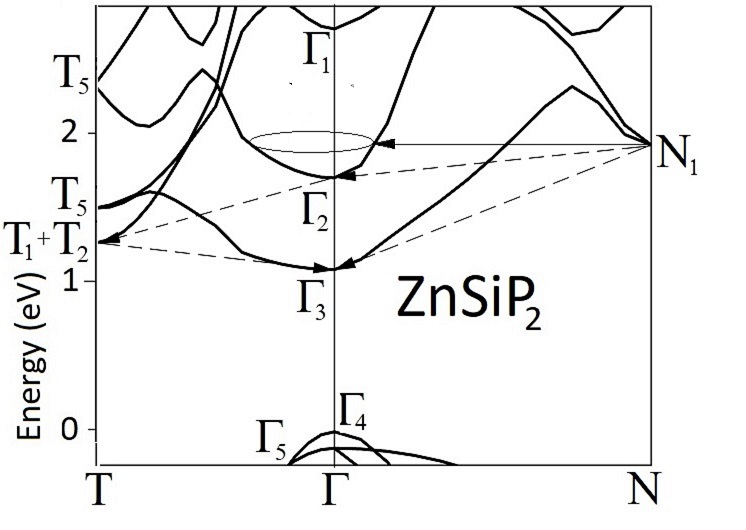}
	\end{minipage}
	\begin{minipage}[h]{0.490\linewidth}
		\includegraphics[width=1\linewidth]{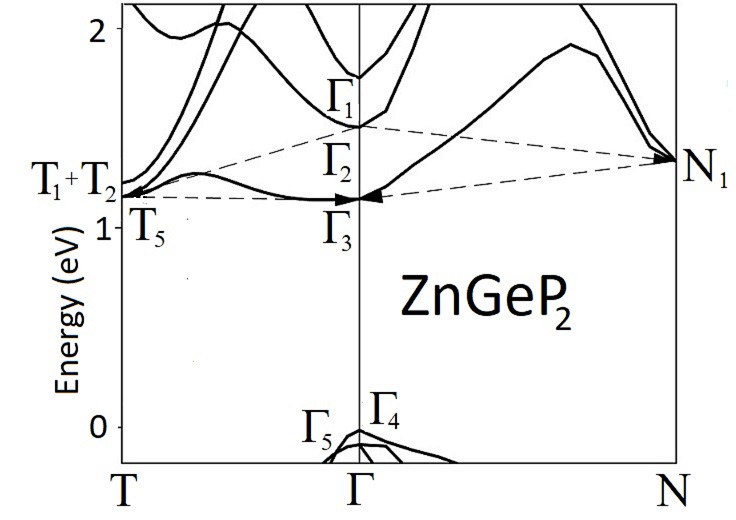}
	\end{minipage}
	\caption{\small{Intervalley transitions in the conduction band of $ZnSiP_2$ and $ZnGeP_2$. Dotted lines represent virtual transitions. As an example, the real transition with energy conservation $ N_1 $ - $ \Gamma_2 $ is shown for $ ZnSiP_2 $ by a solid line.}}
	\label{fig:ZSP_trans}
\end{center}
\end{figure}

 According to our calculation, in $ZnSiP_2$ crystal the level $T_{1c} + T_{2c}$  is the most close to $\Gamma_{3c}$ at the point $T$.  At point $N$ the nearest minimum in energy has symmetry $N_1$.
In $ ZnGeP_2 $  at point $T$ the state $T_ {5c}$ is realized, being very close by energy to $\Gamma_{3c}$. At point $N$, the nearest by energy minimum has the symmetry $N_1$.
Phonons from points $T$ and $N$ are involved into $\Gamma-T$ and $\Gamma-N$ scattering channels correspondingly, their frequencies are doubly degenerate.

The calculated deformation potentials (hereafter in units $eV/\mathring{A}$) in $ZnSiP_2$,(Tabs \ref{Ts:ElphonZSP_1},\ref{Ts:ElphonZSP_2}) varies in range
0.3 $\div$ 3.2 for virtual transitions ($[T_{1c}+T_{2c}]-\Gamma_{3c}$), associated with 24 phonon branches,
0.9 $\div$ 2.0 for  transition ($\Gamma_{2c}-[T_{1c}+T_{2c}]$). In $ZnGiP_2$ deformation potentials varies as
0.9 $\div$ 2.0  for($\Gamma_{2c}-T_{5c}$) and 0.8 $\div$ 3.2 for($T_{5c}-\Gamma_{3c}$)(Tabs \ref{Ts:ElphonZGP_1},\ref{Ts:ElphonZGP_2}).
Note, that these values are close by their values to the deformation potentials in $Ge$
(2.4$\div$4.0 ) \cite{ObukGe}, $Si$(2.51$\div$4.73) \cite{ObukFTTeng} and in $GaP$(3.8) \cite{NikObukhTyut2009}.

Electron-phonon scattering rates (\ref{rates}) for various scattering channels at three temperatures are shown at Tab.\ref{Ts:Elphon_rates}.

\begin{table}[ht]
\footnotesize
	\begin{center}
	\caption{\small{Temperature dependencies of intervalley  electron-phonon scattering rates $R^i_j(T)$
($10^{11} eV/s$) in the conduction bands of $ZnSiP_2$ and $ZnGeP_2$}}\label{Ts:Elphon_rates}
 \begin{tabular}{|c|c|c|c|c|}
  	\hline
   \multicolumn{5}{|c|}{$ZnSiP_2$}\\
   	\hline
  \multicolumn{1}{|c}{$i$}&\multicolumn{1}{|c}{$N_{1c}$}&\multicolumn{1}{|c}{$\Gamma_{2c}$}&
  \multicolumn{1}{|c}{$T_{1c}+T_{2c}$}&\multicolumn{1}{|c|}{$N_{1c}$}\\
  \hline
  \multicolumn{1}{|c}{$j$}&\multicolumn{1}{|c}{$\Gamma_{2c}$}&\multicolumn{1}{|c}{$T_{1c}+T_{2c}$}&
  \multicolumn{1}{|c}{$\Gamma_{3c}$}&\multicolumn{1}{|c|}{$\Gamma_{3c}$}\\
    \hline
 	 $10K$ &6.52  &3.21  &5.59  &7.46 \\
     $77K$ &7.11  &4.00  &5.99  &9.04 \\
    $300K$ &13.61 &10.06 &10.91 &22.09 \\
        	\hline
  \multicolumn{5}{|c|}{$ZnGeP_2$}\\
        	\hline
  \multicolumn{1}{|c}{$i$}&  \multicolumn{1}{|c}{$\Gamma_{2c}$}& \multicolumn{1}{|c}{$\Gamma_{2c}$}&
    \multicolumn{1}{|c}{$N_{1c}$}&  \multicolumn{1}{|c|}{$T_{5c}$}\\
    \hline
  \multicolumn{1}{|c}{$j$}&  \multicolumn{1}{|c}{$N_{1c}$} &  \multicolumn{1}{|c}{$T_{5c}$} &
  \multicolumn{1}{|c}{$\Gamma_{3c}$} &  \multicolumn{1}{|c|}{$\Gamma_{3c}$}\\
    \hline
 	 $10K$ &7.16  &5.99  &7.64  &3.52 \\
     $77K$ &8.36  &6.47  &9.74  &4.01 \\
    $300K$ &18.56 &12.72 &25.04 &9.01 \\
        	\hline

 \end{tabular}
   \end{center}
\end{table}

In accordance with the selection rule for optical transition, electrons in a pseudo-direct-gap semiconductor are excited by light into the overlying minima, bypassing the bottom of the conduction band.
As can be seen from the Tab.\ref{Ts:Elphon_rates}, the inter-valley processes provide a fairly high rate of electron-phonon scattering.
Therefore, the transition of photoexcited electrons to the bottom of the conduction band through intermediate scattering into side valleys can provide an effective relaxation mechanism.

\section{Conclusion.}
Within the framework of a unified, self-consistent quantum-mechanical method, the equilibrium crystal structure of two compounds with the chalcopyrite lattice $ZnSiP_2$ and $ZnGeP_2$ is calculated. The optimized structural parameters are used to calculate the spectra of electrons and phonons, and are in fairly good agreement with experiment and available theoretical calculations. In our calculations, the conduction band minimum in both compounds is realized in the state $\Gamma_{3c}$. The next lowest value of the energy minimum of the conduction band is realized at the point $T$ of Brillouin zone  in the state $T_{1c} + T_{2c}$ in $ZnSiP$ and in the state $T_{5c}$ in $ZnGeP_2$. For the first time, the probabilities of intervalley scattering of electrons in the conduction band of these compounds by phonons are calculated. The intervalley transitions of the most practical interest in both compounds are those of  $\Gamma\longleftrightarrow T$ with the participation of short-wavelength phonons from point $T$. The deformation potentials for
$\Gamma \longleftrightarrow T$ scattering, which involve phonons from the point $T$, and for $\Gamma \longleftrightarrow N$  involving phonons from the $N$ point are calculated by the $DFPT$ method.
All phonons participating in the scattering are doubly degenerated. Values of the calculated intervalley deformation potentials are close to their values in $Si$, $Ge$ and in the binary analog $GaP$.

The knowledge of the probabilities of electron scattering by short-wavelength phonons is necessary in the study of energy relaxation at high excitation levels by external high-energy radiation \cite{TyutZhukEcheniqChulk2015}, as well as in connection with the search of thermoelectric materials with high efficiency \cite{WangObuhTyutPRB_Thermo2011}, sf also recent review article Sjakste et al \cite{Sjakste_JPS} and references therein.

\section*{Acknoledgements.}
The research was supported by Ministry of Education and Science of Russian
Federation, research project No.3.6765.2017/8.9.
\newpage
\section*{Appendix.}
Deformation potentials for intervalley transitions in chalcopyrite crystals were not calculated before.
In $ZnSiP_2$ crystal the level $T_{1c} + T_{2c}$  is the most close to $\Gamma_{3c}$ at the point $T$.
 At point $N$ the nearest in energy to $\Gamma_{2c}$ minimum has the symmetry $N_1$.
Phonon frequencies  both at $T$ and $N$ points are doubly degenerate.
\begin{table}[ht]
\footnotesize
	\begin{center}
		\caption{\small{Parameters of intervalley $\Gamma_{3c}-(T_{1c}+T_{2c})$ scattering of electrons by phonons in the conduction band of $ZnSiP_2$ }}\label{Ts:ElphonZSP_1}
 \begin{tabular}{|c|c|}
 	\hline
 \multicolumn{2}{|c|}{$ZnSiP_2$}\\
 	\hline
 \multicolumn{1}{|c}{$T$-phonon's}&\multicolumn{1}{|c|}{$D$ in units}\\
\multicolumn{1}{|c}{frequency}&\multicolumn{1}{|c|}{$eV/\mathring{A}$}\\
\multicolumn{1}{|c}{ $THz$  }&\multicolumn{1}{|c|}{}\\
   \hline
 	  2.31 &0.3   \\
      2.84 & 0.8  \\
     4.56 & 0.2   \\
     8.32 & 2.5  \\
     9.96 & 0.4  \\
    14.28 &0.8  \\
     14.60 &3.2  \\
     	\hline
 \end{tabular}
   \end{center}
\end{table}
\begin{table}[ht]
\footnotesize
	\begin{center}
\caption{\small{Parameters of intervalley $\Gamma_{2c}-N_{1c}$ scattering of electrons by phonons in the conduction band of $ZnSiP_2$ }}\label{Ts:ElphonZSP_2}
 \begin{tabular}{|c|c|}
 	\hline
 \multicolumn{2}{|c|}{$ZnSiP_2$}\\
 	\hline
  \multicolumn{1}{|c}{$N$-phonon's}&\multicolumn{1}{|c|}{$D$ in units }\\
\multicolumn{1}{|c}{ frequency }&\multicolumn{1}{|c|}{$eV/\mathring{A}$}\\
\multicolumn{1}{|c}{ $THz$  }&\multicolumn{1}{|c|}{}\\
\hline
 	2.26  &  0.6 \\
	2.55  &   0.2 \\
	 3.07 &   0.7\\
	4.90  & 0.8 \\
	6.82  & 1.1  \\
	 8.56 &   1.5\\
	9.87  & 0.3  \\
	  10.25 & 1.8  \\
	  11.34 &  1.9 \\
	  14.15 &   1.8\\
	 14.28 &  2.0 \\
	  14.60 &  0.8 \\
 	\hline
 \end{tabular}
   \end{center}
\end{table}
In $ ZnGeP_2 $  at point $T$ the state $T_ {5c}$ is realized, being very close by energy to $\Gamma_{3c}$ (sf Tab.\ref{Ts:Levels}). At the point $N$, the nearest by energy minimum has the symmetry $N_1$.
\begin{table}[ht]
\footnotesize
	\begin{center}
		\caption{\small{Parameters of intervalley $\Gamma_{3c}-T_{5c}$ scattering of electrons by phonons in the conduction band of $ZnGeP_2$ }}\label{Ts:ElphonZGP_1}
 \begin{tabular}{|c|c|}
 	\hline
 \multicolumn{2}{|c|}{$ZnGeP_2$}\\
 	\hline
 \multicolumn{1}{|c}{$T$-phonon's }&\multicolumn{1}{|c|}{$D$ in units}\\
\multicolumn{1}{|c}{ frequency  }&\multicolumn{1}{|c|}{$eV/\mathring{A}$}\\
\multicolumn{1}{|c}{ $THz$  }&\multicolumn{1}{|c|}{}\\
\hline
  2.56 &0.9  \\
  4.85 & 1.4  \\
    10.31 &2.3   \\
   10.92 & 1.3   \\
   11.52 & 0.9  \\
     	\hline
 \end{tabular}
   \end{center}
\end{table}
\begin{table}[ht]
\footnotesize
\begin{center}
\caption{\small{Parameters of intervalley $\Gamma-N$ scattering of electrons by phonons in the conduction band of $ZnGeP_2$  }}\label{Ts:ElphonZGP_2}
 \begin{tabular}{|c|c|c|}
 	\hline
 \multicolumn{3}{|c|}{$ZnGeP_2$}\\
 	\hline
 \multicolumn{1}{|c}{$N$-phonon's}&\multicolumn{2}{|c|}{$D$ in units }\\
\multicolumn{1}{|c}{ frequency }&\multicolumn{2}{|c|}{$eV/\mathring{A}$}\\
\multicolumn{1}{|c}{ $THz$  }&\multicolumn{2}{|c|}{}\\
\cline{2-3}
\multicolumn{1}{|c}{}&\multicolumn{1}{|c}{Scattering}&\multicolumn{1}{|c|}{Scattering} \\
\multicolumn{1}{|c}{ }&\multicolumn{1}{|c}{$\Gamma_{3c}$ --$N_{1c}$}&\multicolumn{1}{|c|}{ $\Gamma_{2c}$ --$N_{1c}$}\\
    \hline
 	 1.99 & 0.9 &0.7  \\
	 2.26 & 1.2 & 0.9  \\
	 2.79 & 0.7 &0.4  \\
	 3.70 &0.6  & 0.5 \\
	 5.49 & 1.5 & 0.8 \\
	  6.71  & 0.4 &1.8  \\
	 9.81 & 1.9 & 1.2 \\
	 10.06  & 1.4 & 2.1 \\
	  10.87 &1.9  & 2.2 \\
	  11.07 & 1.7 & 1.7 \\
	 11.19 & 1.4 &1.5  \\
	 11.66 &1.0  & 1.3 \\
 	\hline
 \end{tabular}
   \end{center}
\end{table}
For frequencies not specified in Tabs.\ref{Ts:ElphonZSP_1},\ref{Ts:ElphonZGP_1} the deformation potentials vanish by symmetry.
\newpage
\bibliographystyle{elsarticle-num}
\bibliography{Chalk-p}
\end{document}